\documentclass[useAMS]{mn2e}

\usepackage{subfigure}
\usepackage{amssymb}
\usepackage{graphics}
\usepackage{amsmath}
\usepackage{amsfonts}
\usepackage{bm}
\usepackage[dvips]{color}

\usepackage{graphicx}

\newcommand*{\be}{\begin{equation}} 
\newcommand*{\ee}{\end{equation}}
 
\title{Nonlinear electrodynamics and the variation of the fine structure
constant}

\author[Mbelek and Mosquera Cuesta]{Jean Paul Mbelek$^{1,2}$ and Herman J. 
Mosquera Cuesta$^{3,4}$ \\
$^1$Service d'Astrophysique, CEA-Saclay, Paris, France\\
$^2$Department of Physics, Faculty of science, University of Ngaound\'er\'e,
P.O.Box 
454, Ngaound\'er\'e, Cameroon\\  
$^3$\mbox{Instituto de Cosmologia, Relatividade e
Astrof\'{\i}sica (ICRA-BR), Centro Brasileiro de Pesquisas
F\'{\i}sicas (CBPF)}\\ Rua Dr.  Xavier Sigaud 150, 22290-180,
RJ, Brasil :::  hermanjc@cbpf.br \\
$^4$\mbox{ICRANet, Piazzalle della Repubblica 10, I-65100, Pescara, Italy} }
 
\date{\today}

\begin{document}

\maketitle

\label{firstpage}



\begin{abstract}
It has been claimed that during the late time history of our universe, 
the fine structure constant of electromagnetism, $\alpha$, has been 
increasing (Webb et al. 2001; Murphy et al. 2003). The conclusion is 
achieved after looking at the separation between lines of ions
like CIV, MgII, SiII, FeII, among others in the absorption spectra of 
very distant quasars, and comparing them  with their counterparts  
obtained in the laboratory.

However, in the meantime, other teams has claimed either a null result or 
a decreasing $\alpha$ with respect to the cosmic time (Chand et al. 2004;
Levshakov et al. 2004). Also, the current precision of laboratory tests 
does not allow one to either comfort or reject any of these astronomical 
observations. Here we suggest  that as photons are the sidereal messengers, 
a nonlinear electrodynamics (NLED) description of the interaction 
of photons with the weak local background magnetic fields of a gas cloud
absorber around the emitting quasar can reconcile the Chand et al. (2004) and
Levshakov et al. (2004)  findings with the negative variation found by Murphy 
et al. (2001a, 2001b, 2001c, 2001d) and Webb et al. (2001), and also to find
a bridge with the positive variation argued more recently by Levshakov et al. 
(2006a, 2007). We also show that nonlinear electrodynamics photon propagation 
in a vacuum permeated by a background magnetic field presents a full agreement with 
constraints from Oklo natural reactor data. Finally, we show that NLED may render 
a null result only in a narrow range of the local background magnetic field which
should be the case of both the claims by Chand et al.(2004) and by Srianand et al. 
(2004).
\end{abstract}

\begin{keywords}
Line absorptions; variation of alpha; cosmology; nonlinear electromagnetism 
\end{keywords}


\section{Introduction}

The gauge gouplings $\alpha_G, \alpha, \alpha_W$, and so on,
of theories of fundamental  interactions as gravity,
electroweak, strong and GUTs are 
by definition constant! That is, they are supposed not to depend on time!
nor on the spatial location. 
The cosmological variation of fundamental physical constants was first
addressed by Dirac in his theory 
on the {\sl Large Number Hypothesis} (Dirac 1937). The impact of this
study on fundamental research has 
extended from multidimensional to the modern Kaluza-Klein and
brane-world theories. However, if one 
conceives for instance the variation in time of a fundamental parameter
as the gravitational constant, 
then one should be prepared to admit a modification of the standard
model of electroweak and strong 
interactions. Pathways into those directions have been pursued and 
their predictions have been demanding more
precision tests of such ``nouvelle'' theories. In the perspective of 
such potential changes, physicists and astronomers have devised skills 
over more than six decades , methods and techniques in an 
attempt to detect and measure any putative variation of these
fundamental constants.

Each formulation of any of the basic physical theories quoted earlier
has its own way to make it evident the possible space-time or simply 
cosmological evolution of those gauge couplings. Usual tests of the 
validity of the ``constant'' hypothesis
 use either experimental setups in laboratories as well as local geophysical
data or deep-space astronomical observations, all of them having in mind 
to at least put tight bounds on their eventual time evolution. For instance, 
a laboratory setup may allow to compare the rate between atomic clocks
built on different composition $(A, Z)$. Meanwhile, the analysis of absorption
spectra from high redshift quasars is at the kernel of most astrophysical
techniques to ask about any cosmological evolution of $\alpha$. In all these 
cases, the main physical reason for this search is that the {\sl splitting 
ratio } of absorption lines depends on $\alpha$ following the Bethe-Weiz\"acker 
relation

\be
\frac{\Delta \lambda}{\lambda} \simeq (Z \alpha)^2 + {\cal{O}}(\alpha)^4
\; ;
\label{lambda-alpha-dependence}
\ee

where $\Delta \lambda = \lambda_1 - \lambda_2$ is the difference of two
spectral lines of 
wavelength $\lambda_1$ and $\lambda_2$. This splitting ratio implies
that the variation of 
$\alpha$ at redshfit $z$ compared to its local (laboratory value) in
first approximation 
reads

\be
\frac{\Delta \alpha}{\alpha} \simeq \frac{1}{2} \left[
\frac{ \left( \frac{\Delta \lambda}
{\lambda} \right)_z }{ \left(\frac{\Delta \lambda}{\lambda} \right)_0} -
1 \right] \; .
\ee


Although most of the above quoted experimental searches produced null
results (Murphy et al. 2004, Quast et al. 2004; Bahcall et al. 2004; Srianand 
et al. 2004), evidence for the 
time variation of the fine structure constant reported from high redshift
quasar absorption systems (Webb et al. 1999, 2001; Murphy 2001a; Murphy et 
al 2001b,2003; Levshakov et al. 2006a,2007) has recently appeared. In similar 
lines, geophysical methods that use both the natural nuclear reactor
that was active $1.8 \times 10^9$ years ago in Oklo Uranium 
Mine in Gabon (West Africa) \footnote{This phenomenon was discovered by 
the french atomic energy commission (Commissariat \`a l'Energie Atomique) in
1972, and consists in an 
abnormally low relative concentration of the isotopes of Samarium - the 
isotopic ratio of the $^{149}$Sm and $^{147}$Sm isotopes. The standard 
isotopic ratio is $\simeq$ 0.9, whereas the actual measurement gave 
$\simeq$ 0.02. It is as if $^{149}$Sm were depleted due to neutron capture
of thermal neutrons when the natural reactor was active. To understand 
this anomaly in the isotopic ratio one can invoke a different value of 
the capture resonance energy at the time of the reaction. This threshold 
energy depends on $\alpha$, as already shown by the Bethe-Weiz\"acker 
formula, quoted above in Eq.(1).}, and the analysis of natural long-lived $\beta$ 
decayers in geological minerals and meteorites showed no variation 
(Olive et al. 2004). Let us emphasize that the latter data as well as the Oklo
one are out of the cosmological context.

Bounds on the variation of $\alpha$ in the early universe can
be obtained using data from the 
Cosmic Microwave Background (CMB) radiation and from the abundances of
light elements. Although these 
bounds are not as stringent as the mentioned above, they are important
because they refer to a different
cosmological era (see Mosquera, Scoccola, Landau \& Vucetich 2007) for 
an account of 
bounds on $\alpha$ from the early universe based on abundances of D,
$^3{\rm He}$, $^4{\rm He}$ and 
$^7{\rm Li}$, and also see the recent review by Garc\'{\i}a-Berro, Isern 
and Kubyshin 2007).

Just very recently, a very innovative experimental proposal in which the 
effects of variation in the value of the fine structure constant at extremely 
high redshfits, $z_{\rm rec} \lesssim z \lesssim 30$, on the absorption of the 
cosmic microwave background radiation (CMB) at the 21 cm of the hyperfine transition 
of neutral atomic hydrogen have been investigated (Khatri and Wandelt 2007).
It is shown there that the 21 cm signal is very sensitive to variations of
$\alpha$, so that a change in the value of $\alpha$ by 1\% changes the mean 
brightness temperature decrement of the CMB due to 21 cm absorption by $>5\%$ 
over the redshift range $z<50$. That paper also demonstrates that the signal 
from the 21 cm absorption by neutral hydrogen between redshifts of 30 and 200 is 
extremely sensitive to the value of $\alpha$ during that epoch. By assuming that
the technological hurdles for detecting this absorption signal in the long
wavelength radio band can be overcome, the change of $\alpha$ from the era of 
redshift $\sim$100 can be constrained to 1 part in $10^9$ (Wandelt 2007). 

Our objective 
in this paper is to call to the attention of workers in the field that nonlinear 
electrodynamics can help to understand the current controversial observations indicating 
a putative variation of alpha. In a future communication we plan to address each individual 
source that have been observed by all of the research groups cited in this paper to
confront each of them with the predictions of the NLED theory here brought into the 
discussion.


\section{Quasar absorption spectra and the controversy on variation of
$\alpha$}


Since long ago astronomers  realized that ideal laboratories to  search
for a potential signature 
of cosmic evolution of the fine structure constant are high-redshift
quasars whose spectra present 
absorption resonance lines of several alkaline ions  like CIV, MgII,
FeII, including the SiIV 
doublet system that is observed from a source at redshift between $2.5 \leq z
\leq 3.33$ (see Mosquera et al. 2007, and references therein). Also 
OIII emission lines have been 
observed (Bahcall et al. 2004). Other authors used transitions between
species with far different atomic 
masses from which they obtained a single data consistent with time
varying of $\alpha$ at redshift 
$0.5 \leq z \leq 3.5$ (Webb et al. 1999; Murphy et al. 2001a, 2003). 
Notwithstanding, no variation was found 
from other recent independent analysis of similar observations
(Quast et al. 2004,; Srianand et al. 2004; Grupe et al. 2005). 
Meanwhile, Levshakov et al. introduced a different procedure in which a
high resolution spectrograph
is used to observe pairs of FeII lines during individual exposures
(Levshakov et al. 2005). No variation 
of $\alpha$ for redshifts $z = 1.15$ and $z = 1.839$ was found. Very
recently, however, a reanalysis 
of the spectrum of the quasar Q 1101-264 exhibit signatures of
variability with a confidence level 
of $1 \sigma$ (Levshakov et al. 2007). 

From another side, techniques that compare molecular and radio lines 
provided more stringent constraints (Murphy et al. 2003). Also bounds on
cosmological evolution of $\alpha$ at redshfit $z = 0.2467$ obtained from  
satellite observations at wavelength $\lambda = 18$ cm for the $OH$ conjugate 
lines have been reported (Darling et al. 2004). Moreover, Kanekar et al. (2005) 
compared the redshifts of the HI and OH main absorption lines of the different 
components in an absorber source at $z = 0.765$, and also of the lens toward 
the object B0218+357 at $z = 0.685$, to place stringent constraints on changes 
in the parameter $F$ defined as $F = g_p \left( {\frac{\alpha^2}{\mu}}
\right)^{1.57}$. Finally, a full analysis of the bounds on cosmological variation 
of $\alpha$ obtained by comparing the optical and radio redshifts 
is presented by Mosquera et al. (2007) based on data from
Refs.(Wolfe et al. 1976; Spinrad and Mackee 1979; Cowie and Songaila 
1995; Tzanaris, et al. 2007).

In summary, the controversy is still lively pushed ahead since the moment 
in which Webb et al. (2001) and Murphy et al. (2001a, 2001b, 2001c, 2001d) 
claimed a negative variation of the fine structure constant as a function 
of the redshift. In the meantime,  Levshakov et al.(2006a) have 
found a positive variation of $\alpha$  
between the two redshifts quoted above. The null results of Chand et al. 
(2004) and Srianand et al. (2004) could appear as an intermediate case 
making the transition between the former two\footnote{It has been pointed 
out that this discrepancy comes out only because those authors have chosen 
to use solar abundances of Mg$^{25,26}$ (Ashenfelter et al. 2004).} but it 
has been challenged recently by Murphy et al. (2007e), who have pointed 
out a large number of errors in the statistical analysis performed by Chand 
et al. (2004). Indeed, Murphy et al. (2007e) show that when these are corrected 
for one finds $\Delta 
\alpha/\alpha = -(0.64 \pm 0.36) \times 10^{-5}$, which is a 6-fold larger 
uncertainty than that quoted by Chand et al. (2004). In passing, such a 
value of $\Delta \alpha/\alpha$ is consistent with the value found by Webb 
et al.(2001).

In view of those
discrepant data one may wonder whether some basic systematic effects, 
yet not accounted for, may be involved in the aforementioned 
observations. As a possible pathway to unveiling these effects let us
recall that our knowledge in astrophysics and cosmology comes out mostly 
from the information gathered from electromagnetic (EM) waves (as far 
as gravitational waves  start to be detected in the near future). 
In this perspective, it has been proved that electrodynamics in a vacuum
is subject to nonlinear effects (Burke et al. 1997). Hence, it is 
legitimate to address the question of the possible variation of the 
fine structure constant, $\alpha$, within the framework of NLED. It is 
interesting to notice that based on Maxwell electromagnetic theory, M. 
T. Murphy et al. considered large magnetic fields as a potential cause 
of systematic errors in their measurements of $\Delta \alpha/\alpha$ 
(see Murphy et al. 2001, Section 2.6). They concluded that the intra-cluster
magnetic field strengths are nine orders of magnitude below the strength 
required to cause substantial effects. As we show below, 
the latter conclusion can be reversed by considering the NLED Lagrangian
density proposed by Novello, P\'erez-Bergliaffa and Salim (2004), which 
was introduced as an alternative to dark energy to provide an explanation 
of the SNIa observation-inspired interpretation of a late-time 
acceleration of the expansion of the universe.


\section{NLED: Photon vacuum nonlinear interaction and the variation of
the fine structure 
constant}

A nonlinear Lagrangian is one for which the second, or even higher,
derivatives of the Lagrangian with respect to the Maxwell invariant 
$F = F_{\mu \nu} F^{\mu \nu}$ 
is not null! That is, $L_{FF} \neq 0$, where $L_F= \partial 
L/\partial F$ and $L_{FF}= \partial^2L/\partial F^2$. This is clearly 
not the case for Maxwell's theory. Therefore, one way to guide ourselves 
to build a NLED Lagrangean able to account for the unavoidable nonlinear 
interaction of photons from distant quasars with the intergalactic background 
EM field is to realize that those background fields are extremely weak,
so that in order to have any 
practical influence on the photon dynamics the Lagrangian should depend 
on the $F$ field in a nontrivial fashion. 
This is our approach in what follows. We first start with the dynamics
of the photon nonlinear 
propagation in a vacuum. Then we will focus on the structure of that
specific Lagrangian, and its 
dynamical consequences.

\subsection{Photon dynamics in NLED}

In this Section we investigate the effects of nonlinearities in the
evolution of EM waves in a vacuum, where the waves are described 
onwards as the surface of discontinuity of the EM field. Extremizing 
the Lagrangian $L(F)$, with $F(A_\mu)$, with respect to the potentials 
$A_{\mu}$ yields the following field equation (Plebanski 1970)

\be 
\nabla_{\nu} (L_{F}F^{\mu\nu} ) = 0\label{eq60} , 
\ee

where $\nabla_\nu $ defines the covariant derivative. Besides this, 
we have the cyclic identity

\be
\nabla_{\nu} F^{*\mu\nu} = 0 \hskip 0.3 truecm \Leftrightarrow 
\hskip 0.3 truecm F_{\mu\nu|\alpha} + F_{\alpha\mu|\nu} + 
F_{\nu\alpha|\mu} = 0\; . 
\label{eq62}
\ee

Taking the discontinuities of the field equation we get
\footnote{ Following Hadamard (1903), the surface of 
discontinuity of the EM field is denoted by $\Sigma$. The 
field is continuous when crossing $\Sigma$, while its first derivative
presents a finite 
discontinuity. These properties are specified as follows: $\left[F_{\mu
\nu} \right]_{\Sigma} 
= 0\; ,$ \hskip 0.3 truecm $\left[F_{\mu\nu|\lambda}\right]_{\Sigma} =
f_{\mu\nu} k_\lambda\; 
\protect \label{eq14} \;$, where the symbol $\left[F_{\mu
\nu}\right]_{\Sigma} = \lim_{\delta 
\to 0^+} \left(J|_{\Sigma + \delta}-J|_{\Sigma - \delta}\right)$
represents the discontinuity 
of the arbitrary function $J$ through the surface $\Sigma$. The tensor
$f_{\mu\nu}$ is called 
the discontinuity of the field,  $k_{\lambda} = \partial_{\lambda}
\Sigma $ is the propagation 
vector, and the symbols "$_|$" and "$_{||}$" stand for partial and
covariant derivatives.} 

\be 
L_{F} f^{\; \; \mu}_{\lambda} k^\lambda + 2L_{FF}F^{\alpha\beta} 
f_{\alpha\beta} F^{\mu\lambda} k_{\lambda} = 0 \; ,
\label{j1} 
\ee

which together with the discontinuity of the Bianchi identity yields

\be
f_{\alpha\beta}k_{\gamma} + f_{\gamma\alpha}k_{\beta} + 
f_{\beta \gamma} k_{\alpha} = 0 .
\ee

In order to obtain a scalar relation we contract this equation
with $ k^{\gamma}F^{\alpha\beta} \label{eq25}$, resulting

\be
(F^{\alpha\beta}f_{\alpha\beta} g^{\mu\nu} + 2F^{\mu\lambda} 
f_{\lambda}^{\; \; \nu})k_{\mu} k_{\nu}=0 \; .
\label{j2}
\ee

It is straightforward to see that here we find two distinct solutions: a) when
$F^{\alpha\beta} f_{\alpha\beta}=0$, case in which such mode propagates 
along standard null geodesics, and b) when 
$F^{\alpha\beta} f_{\alpha\beta}=\chi$. In the case in which $\chi$ does
not vanish 
we obtain from equations (\ref{j1}) and (\ref{j2}), the propagation
equation for the 
field discontinuities being given by (Novello et al. 2004)

\be
\underbrace{ \left(g^{\mu\nu} - 4\frac{L_{FF}}{L_{F}} F^{\mu\alpha} 
F_{\alpha}^{\; \; \nu}\right) }_{\rm effective \; metric} k_{\mu}k_{\nu} = 0 \;
.
\label{63}
\ee

This equation proves that photons propagate following a geodesic that is
not that one of the background space-time described by $g^{\mu\nu}$. Rather,
they follow the {\sl effective metric } given by Eq.(\ref{63}).

If one now takes the $x^a$ derivative of this expression, we can easily
obtain (Mosquera Cuesta, de Freitas Pacheco and Salim 2006; Mosquera Cuesta 
and Salim 2004, Mosquera Cuesta and Salim 2004A)

\be
k_{\alpha||\nu} k^{\nu} = 4 \left(\frac{L_{FF}}{L_{F}}F^{\mu\beta}
F_{\beta}^{\; \; \nu} ~k_{\mu} k_{\nu}\right)_{|\alpha}.
\label{kuknu}
\ee

This expression shows that the nonlinear Lagrangian introduces a term
acting as a force that accelerates the photon along its path. It is 
therefore essential to investigate what are the effects of this 
peculiar prediction. The occurrence of this phenomenon over cosmological
distance scales may have a nonnegligible effect on the physical properties 
that are abscribed to a given source from its astronomical observables. One 
example of this is the cosmological redshift (Mosquera Cuesta, Salim and 
Novello 2007). Since the photon ought to travel very long distances from 
cosmic sources until be detected on Earth, then its interaction with the 
background intergalactic electromagnetic fields should modify the putative 
(nominal) value of the redshift, or equivalently, the actual luminosity distance, 
associated to its emitting source, compared to the proper distance computed 
in a standard fashion in the context of a particular cosmology. In similar lines, 
we show next that in the case of EM radiation coming from far distance radio-galaxies 
and quasars the interaction of the photon with local intergalactic background 
EM fields may significantly modify the actual position of a particular
absorption line, from which a potential variation of $\alpha$ can be estimated. 
In this way, an observer on Earth is prone to say that effectively $\alpha$ has 
changed for this particular observation. But, how exactly does it change?


\subsection{NLED and cosmological variation of $\alpha$}

In order to investigate whether the photon nonlinear interaction with
background fields over large distances does affect the position of 
absorption lines, next we present the Lagrangian formulation of 
this NLED theory. 

{

\subsubsection{A motivation to look for nonlinear electrodynamics effects 
in cosmology} 

In Novello et al. (2004) several general properties of nonlinear electrodynamics 
in cosmology were reviewed by assuming that the action for the electromagnetic 
field is that of Maxwell with an extra term, namely \footnote{Notice that this 
Lagrangian is gauge invariant, and that hence charge conservation is guaranteed 
in this theory.}

\be
S = \int \sqrt{-g} \left( - \frac F 4 + \frac \gamma F \right) d^4x \; ,
\label{action}
\ee

where $F\equiv F_{\mu\nu}F^{\mu\nu}$. 

Physical motivations for bringing in this 
theory have been provided in Novello et al. (2004).  Besides, an equally unavoidable 
motivation comes from the introduction of both the Heisenberg-Euler and Born-Infeld 
nonlinear electrodynamics, which are valid in the regime of extremely high magnetic 
field strengths. Those studies constituted the first demonstration that a Quantum 
Electrodynamics (QED) description of the atomic world would be neccessary, at least; 
at the one-loop level to overcome the {\it ultraviolet catastrophe} \footnote{Also 
called the Rayleigh-Jeans catastrophe, it was a prediction of early 20th century 
classical physics that an ideal black body (BB) at thermal equilibrium will emit 
radiation with infinite power. The term {\it ultraviolet catastrophe} was first 
used in 1911 by Paul Ehrenfest, although the concept goes back to 1905 study of 
BB by Planck; the 
word {\it ultraviolet} refers to the fact that the problem appears in the short 
wavelength region of the electromagnetic spectrum. Since the first appearance 
of the term, it has also been used for other predictions of a similar nature, 
as in quantum electrodynamics and such cases as ultraviolet divergence in Quantum 
Field Theory (QFT).}. Both 
theories have been extensively investigated in the literature (see for instance 
Mosquera Cuesta and Salim 2004; Mosquera Cuesta and Salim 2004A; Mosquera Cuesta, 
de Freitas Pacheco and Salim 2006, and the long list of references therein). Since 
in nature non only such very strong magnetic fields exist, then it appears to be 
promising to investigate also those super weak field frontiers that should outcome
over intergalactic distances, i. e., over the space encircling the absorption line
emitting galaxy.

Regarding this phenomenological Lagrangian in Eq.(\ref{action}), at first, one 
notices that for high values of the field $F$, the dynamics resembles Maxwell's 
one except for small corrections associate to the parameter $\gamma$, while at 
low $B$-field strengths, i. e., $F \rightarrow 0$, it is the $1/F$ term that dominates.
\footnote{That is why we did not address the description of the standard physical 
picture called for when one considers the electromagnetic interaction inside a 
hydrogen atom, for instance. At such a distance scale the electric fields produced 
by both the electron shell and the nucleus overrun even the Maxwell limit, so that 
one is forced to invoke Heisenberg-Euler or Born-Infeld nonlinear electrodynamics
to more accurately figure out what is electrically going on at that level, rather 
than invoking this $F + F^{-1}$ theory. {\it Therefore, the correction can be safely 
neglected also in the case of the energy spectrum of the hydrogen atom}. We thank 
Prof. E. Elbaz, Universite de Lyon, France, for this important discussion in a 
private communication.} Clearly, 
this last term should dramatically affect the photon-$\vec{B}$ field interaction 
over the intergalactic space. In this respect, the consistency of this theory with 
observations, including the recovery of the well-stablished Coulomb law, was 
shown by Novello et al. (2004) using the cosmic microwave radiation bound, and 
also after discussing the anomaly in the dynamics of Pioneer 10 spacecraft 
by Mbelek et al. (2006). Both analysis provide small enough values for the 
coupling constant $\gamma$. 

Therefore, the electromagnetic (EM) field described by Eq.(\ref{action}) can be 
taken as source in Einstein equations, to obtain a toy model for the evolution of 
the universe which displays accelerate expansion caused when the nonlinear EM term 
takes over the term describing other matter fields. This NLED theory yields ordinary 
radiation plus a dark energy component with $w < -1$ (phantom-like dynamics). 
Introducing the notation \footnote{Due to the isotropy of the spatial sections 
of the Friedman-Robertson-Walker (FRW) model, an average procedure is needed if 
electromagnetic fields are to act as a source of gravity (Tolman and Ehrenfest ). 
Thus a volumetric spatial average of a quantity $X$ at the time $t$ by $\langle X 
\rangle_{|_V} \equiv \lim_{V\rightarrow V_0} \frac 1 V \int X \sqrt{-g}\;d^3x$, 
where $V = \int \sqrt{-g} \;d^3x$ , and $V_0$ is a sufficiently large time-dependent 
three-volume. (Here the metric sign convention $(+---)$ applies).}, the EM field can 
act as a source for the FRW model if $\langle E_i \rangle_{|_V} =0, \; \langle B_i 
\rangle_{|_V} =0,\; \langle {E_i B_j} \rangle_{|_V} = 0$, $ \langle {E_iE_j} 
\rangle_{|_V} = - \frac 1 3 E^2 g_{ij}$, and $\; \langle {B_iB_j} \rangle_{|_V} = 
-\frac 1 3 B^2 g_{ij}$.\footnote{Let us remark that since we are assuming that 
$\langle {B}_i \rangle_{|_V} = 0$, the background magnetic fields induce no 
directional effects in the sky, in accordance with the symmetries of the standard 
cosmological model.} When these conditions are fulfilled, a general nonlinear 
Lagrangian $L(F)$ yields the energy-momentum tensor ($L_F = {dL}/{dF}, \;\; L_{FF} 
= {d^2L}/{dF^2}$)\footnote{Under the same assumptions, the EM field associate to 
Maxwell Lagrangian generates the stress-energy tensor defined by Eq.(\ref{tmunu}) 
but now $ \rho = 3 p = \frac{1}{2} (E^2 + B^2)$.}  

\begin{eqnarray}
& \langle {T}_{\mu\nu} \rangle_{|_V}  =  (\rho + p) v_\mu v_\nu - p\; g_{\mu\nu} 
\;, & \label{tmunu} \\ 
& \rho  =  -L - 4E^2 L_F , \;\;\;\;\; p =  L + \frac 4 3 (E^2-2B^2) L_F \; , 
& \nonumber
\end{eqnarray}

Hence, when there is only a magnetic field, the fluid can be thought of as 
composed of ordinary radiation with $p_{1}= \frac 1 3\; \rho_{1}$ and of another fluid 
with EOS $p_{2} = -\frac 7 3 \;\rho_{2}$. It is precisely this component with negative 
pressure that may drive accelerate expansion throughout the Friedmann equations.  
} 

It is important to notice that this theory has been successfully applied to both the 
study of the late-time cosmic acceleration inferred from supernovae type Ia observations 
(Novello et al. 2004), and also to explain the nontrivial phenomenon known as the 
``Anomaly of the Pioneer Spacecraft'' (Mbelek et al. 2006), a problem that has been 
around over nearly two decades. In a separate communication (Mosquera Cuesta, Salim
and Novello 2007) we show how the cosmological redshift is modified in virtue of 
this unaccounted effect from nonlinear propagation of EM radiation from cosmic 
sources. The consequences of this effect appear to be dramatic for current 
astronomical investigations. Finally, we also advance that the Compton Effect 
itself is radically alterated by this unaccounted NLED effect (Mbelek and 
Mosquera Cuesta 2008).

Turning back to the general NLED Lagrangian $L = L(F)$, one notices that 
in the presence of sources the field equation (\ref{eq60}) becomes  (see 
Novello et al. (2004), Eq.(18))

\be
\nabla_{\nu} (- 4 \; \epsilon_0 \; L_F \; F^{\mu \nu} ) = J^\mu/c^2\; ,
\label{eq.2}
\ee

where $F = F_{\mu \nu} F^{\mu \nu} = 2(B^2c^2 - E^2)$. Hence, the definition of
the NLED effective permittivity
in a vacuum reads

\be
\epsilon_{0\;{\rm NLED}} = - 4 \; \epsilon_0 \; L_F  \; ,
\label{eq.3}
\ee

and consequently the definition of the NLED effective fine structure
constant

\be
\alpha_{\rm NLED}  = \alpha (- 4 \; L_F)^{-1} \; ,
\label{eq.4}
\ee

where $\alpha = $ $e^2/(4\pi\;\epsilon_0\;\hbar\;c)$ with $e$ the electron charge,
$\hbar$ the Planck constant and $c$ the speed of light. Once again, from this 
expression it becomes clear that Maxwell's theory cannot explain any variation 
of alpha.

If one introduces the particular Lagrangian density\footnote{In order to prevent
the Lagrangian under consideration from being singular for $F = 0$ (case of pure
radiation) and for some value in the case of the electric field solely, one may consider a more general expression like $L = - \frac{1}{4}
~F ~+ ~\frac{\gamma}{\sqrt{F^{2} ~+ ~\beta^{2}}}$ which is the sum of the Maxwell Lagrangian and the inverse of the Born-Infeld Lagrangian, and the Latter is known to be non-singular and stable. Thus, one recovers the Lagrangian $L = -
\frac{1}{4} ~F ~+ ~\frac{\gamma}{F}$ for $\mid F \mid \gg \beta > 0$. Notice that
the limiting case $F \rightarrow 0$ ($\mid F \mid \ll \beta$) is equivalent
to Maxwell theory plus a cosmological constant term $\Lambda = \gamma/2\beta$; 
see note [27] in the reference {\it astro-ph/0608538v1} } 

\be
L = - 1/4 \; F + \gamma/F, \; \hskip 0.5 truecm {\rm where} \; \; \;
\gamma = - (B_1 c)^4 \; ,
\label{eq.5}
\ee

(being $\gamma$ a universal constant, and $B_1$ a critical nonzero limiting 
value for the magnetic field strength) equation (\ref{eq.5}) can be rewritten as 
(see Novello et al. (2004), Eqs.(12))

\be
[ \left\{1 + (4\; \gamma/F^2 ) \right\} \; F^{\mu \nu} ]_{\; ;\nu } =
0 .
\label{eq.6}
\ee

As in previous works (Novello et al. 2004; Mbelek et al. 2006), we 
consider hereafter the case with an average magnetic field and with 
a null mean electric field. Therefore, equation (\ref{eq.4}) yields

\be
\alpha_{\rm NLED}  = \alpha \;[ 1 - (B_1/B)^4]^{-1} \; .
\label{eq.7}
\ee

Once again, were the NLED effect null, the $B_1$ field would be zero.
Clearly, since $L_F < 0$ (positive energy condition) and hence $B > B_1
$, equation (\ref{eq.7})  leads to a positive departure of $\alpha_{\rm
NLED}$ from the unperturbed fine structure constant $\alpha$.

\section{Encompassing both local NLED and cosmological effects on
$\alpha$ 
measurements}

Unification theories which are generalization of general relativity
(GGR) like 
superstring theory, scalar-tensor or multidimensional gravitational
theories 
predict the variation of the effective fine structure constant
$\alpha_{\rm 
GGR}$ with respect to the redshift $z$ (or the cosmic time) in the
cosmological 
context (Garc\'{\i}a-Berro, Isern and Kubyshin 
2007; Mbelek and  Lachieze-Rey 2003; Gardner 2003; Sandvik, Barrow and 
Magueijo 2002). Hence, the observed fine structure constant,
$\alpha_{\rm obs}$, might be understood as resulting from a combination 
of both local NLED and cosmological GGR effects, so that

\be
\alpha_{\rm obs}(z)  = \alpha_{\rm GGR}(z) \;[ 1 - (B_1/B)^4]^{-1} \; .
\label{eq.8}
\ee

We show below that this approach can help to reconcile the
negative variation claimed by Webb et al. (2001) and Murphy et al.
(2001a,2001b,2001c), with the recent positive variation 
found by Levshakov et al. (2007) which amounts

\be
\langle \Delta\alpha_{\rm obs}/\alpha \rangle = (0.543 \pm 0.252)~10^{-5}
\ee

between the redshifts $z_1  = 1.15$ and $z_2 = 1.84$ (Levshakov et al.
2006a). Indeed, on Earth, at present ($z = 0 $), in the best laboratory 
conditions the residual magnetic fields (including the geomagnetic field) 
are such that 

\be
B_{\rm lab} \gg B_1, \hskip 0.5truecm {\rm say}, \hskip 0.5truecm B_{\rm
lab}
\gg 1~\mu{\rm G}
\ee

so that 

\be
\alpha_{\rm lab} = \alpha .
\ee

Thus, by setting 

\be
\Delta\alpha_{\rm GGR}(z) = \alpha_{\rm GGR}(z) - \alpha , \hskip
0.2truecm {\rm and}
\hskip 0.2truecm \Delta\alpha_{\rm obs}(z) = \alpha_{\rm obs}(z) -
\alpha ,
\ee

one obtains

\be
\frac{\Delta\alpha_{\rm obs}(z)}{\alpha} = \frac{(\Delta\alpha_{\rm GGR}
(z)/\alpha) + (B_1/B)^4}{1 - (B_1/B)^4} .
\label{eq.9}
\ee 

Averaging over the redshift range of a given sample of absorbers,
relation (\ref{eq.9}) above yields

\be
\langle \frac{\Delta\alpha_{\rm obs}}{\alpha} \rangle =  \frac{\langle 
\Delta \alpha_{\rm GGR}/\alpha \rangle + (B_1/B)^4}{1 - (B_1/B)^4}.
\label{eq.10}
\ee

We will use the relation $B = 6 (n(HI)/$cm$^{-3})^{0.2}~\mu$G 
proposed by J. P. Vall\'ee for the magnetic field strength within 
intergalactic gas clouds (Vall\'ee 2004). This way one finds 

\be
0.4~\mu{\rm G} \leq B < 1.5~\mu{\rm G}
\label{le-champ-intensite}
\ee

for the sample of Murphy, Webb et al. \footnote{see Murphy (2001c)
relations 
$$
N(HI) \sim \frac{N(H)}{1000} \hskip 0.2truecm {\rm and} \hskip 0.2truecm
-~3 < \log_{10} \left({\frac{n_{H}}{cm^{-3}}} \right) < 0 \; ,
$$
in Section 4.1.2 and Table-1 in Section 4.1.3.}. This result implies
that $B \gg B_1$,
and therefore, it also means that the NLED contribution is quite
negligible for this sample 
given the present estimate $B_1 =  \left(0.008 \pm  0.002 \right)~\mu$G
(Mbelek et al. 2006). Thus, we 
conclude from the measurements of Webb et al. and Murphy et al. that 

\be
\langle \frac{\Delta\alpha_{\rm GGR}}{\alpha} \rangle = (- ~0.543 \pm
0.116 )~10^{-5} < 0
\label{general-alpha-variation}
\ee

over the redshift range $0.2 < z < 3.7$ (Murphy et al. 2003). Therefore,
relation (\ref{eq.10}) can be rewritten as

\be
\langle \frac{\Delta\alpha_{\rm obs}}{\alpha} \rangle =  - \;\frac{(B/B_{\rm
c})^4 - 1}{(B/B_1)^4 - 1},
\label{eq.11}
\ee

where we have set $B_{\rm c} = B_1 (\;- \langle \frac{\Delta\alpha_{\rm
GGR}}{\alpha} \rangle )^{-1/4} = (0.165 \pm 0.033)~\mu$G.

Thus, comparing the magnitude $B$ of the mean magnetic field within an 
intergalactic gas cloud absorber to $B_{\rm c}$, one of the three
following conclusions may be reached:

\vskip 0.5 truecm

1) no variation of $\alpha_{\rm obs}$ should be observed from any sample
of absorbing intergalactic gas cloud such that $B \simeq B_{\rm c}$. This 
could be the case of Chand et al. (2004), Srianand et al. (2004) samples, 
but see Murphy et al. (2007e) quoted above and the answer of Srianand et 
al. (2007),

\vskip 0.5 truecm

2) a negative variation of $\alpha_{\rm obs}$ should be observed from
any sample of absorbing intergalactic gas cloud such that $B > B_{c}$ 
(case of Murphy, Webb et al. samples), 

\vskip 0.5 truecm         

3) a positive variation of $\alpha_{\rm obs}$ should be observed from
any sample of absorbing intergalactic gas cloud such that $B < B_{c}$ 
(case of Levshakov et al. sample).

Meanwhile, notice that the HI column densities are provided neither
by Srianand et al., Chand et al. nor by Levshakov et al. for their studies on the
cosmological variation of the fine structure constant. However, Srianand et al.
pointed out that they have avoided sub-Damped Lyman Alpha systems, i. e., $N(HI)
\geq 10^{19}$ cm$^{-2}$ (Srianand et al. 2004). Nevertheless, Boksenberg and
Snijders (1981) derived limits to the neutral hydrogen column density of

\be 
5~10^{17}~{\rm cm}^{2} < N(H) \leq 2~10^{19}~{\rm cm}^{2}, 
\ee

from their observations of the $z_{abs} = 1.8387$ richest absorpsion system
identified in the optical region towards Q 1101-264.

Hence, on account that $N(H{I}) \sim N(H)/1000$,

\be 
\frac{1}{2}~10^{15}~{\rm cm}^{2} < N(HI) \leq \frac{1}{5}~10^{17}~{\rm cm}^{2}, 
\ee

which implies 

\be
0.069  < \frac{B}{\mu{\rm G}} < 0.144 \hskip 0.2truecm 
\label{b-field-limit}
\ee

in accordance with the magnitude of the magnetic field

\be
0.075 < \frac{B}{\mu{\rm G}} < 0.227 \hskip 0.2truecm {\rm and} \hskip
0.2truecm  1.1~10^{15} < \frac{N(HI)}{{\rm cm}^{-2}}  < 2.6~10^{17}
\ee

derived from relation (\ref{eq.11}) and the new measurement of $<\; \Delta\alpha_{\rm
obs}/\alpha \;>$ by Levshakov et al. (2007). We have estimated the distances 
of the gas cloud absorbers from the Hubble law by using $H_{0} = 68$~km~s$^{-1}$
~Mpc$^{-1}$. We emphasize that the estimates of $B$ in relations 
(\ref{le-champ-intensite}) and (\ref{b-field-limit}) 
are consistent with the avalaible astrophysical data on the magnetic field 
strength within intergalactic gas clouds.

\section{Conclusion}

By using a nonlinear eletrodynamics theory in which the effective
Lagrangian is the first order
approximation (after Maxwell's term) of a polynomial series of inverse
powers of the electromagnetic 
invariant quantity $F$, we have presented a consistent explanation of
the controversial results 
regarding a hypotetical variation of the fine structure constant $\alpha
$ since the recombination 
era.

In these lines, one can state that the large set of observations of
quasar absorption systems
are mapping the structure of the intergalactic
magnetic field in several
directions between the Earth and absorbers in the sky, in addition to the
expansion of the universe as a whole. In other words,
the fact that the 
ballpark of the observations seem to be controversial could be
interpreted as an indication that 
the strength of the local magnetic field in each of the observed systems
is most likely different
from each other. Hence, any conclusive statement on the actual
cosmological evolution of the fine
structure constant $\alpha$ rests on a better understanding of the
intergalactic magnetic field 
strength and structure over the whole sky, and also on more accurate
measurements of the relative 
magnitude of the fine splitting between resonance absorption lines from
far away quasars.


{ \textbf{Acknowledgements:}{\; H.J.M.C. thank Prof. R. Ruffini and the ICRANet 
Coordinating Centre, Pescara, Italy, for hospitality during the final
preparation 
of this paper.} }


\end{document}